\documentclass[english, pra, showpacs, onecolumn,
reprint,
superscriptaddress
]{revtex4-2}
\usepackage[dvipsnames]{xcolor}
\usepackage{babel}
\usepackage{verbatim}
\usepackage{bm}
\usepackage{amsmath}
\usepackage{amssymb}
\usepackage{graphicx}
\usepackage{tabularx}
\usepackage{multirow}
\usepackage{booktabs}
\usepackage{colortbl}
\usepackage{float, extpfeil}
\usepackage{placeins}
\usepackage{makecell}

\usepackage{lineno}  
\newcommand*\patchAmsMathEnvironmentsForLineno[1]{%
  \expandafter\let\csname old#1\expandafter\endcsname\csname #1\endcsname
  \expandafter\let\csname oldend#1\expandafter\endcsname\csname end#1\endcsname
  \renewenvironment{#1}%
     {\linenomath\csname old#1\endcsname}%
     {\csname oldend#1\endcsname\endlinenomath}}%
\newcommand*\patchBothAmsMathEnvironmentsForLineno[1]{%
  \patchAmsMathEnvironmentsForLineno{#1}%
  \patchAmsMathEnvironmentsForLineno{#1*}}%
\AtBeginDocument{%
  \patchBothAmsMathEnvironmentsForLineno{equation}%
  \patchBothAmsMathEnvironmentsForLineno{align}%
  \patchBothAmsMathEnvironmentsForLineno{flalign}%
  \patchBothAmsMathEnvironmentsForLineno{alignat}%
  \patchBothAmsMathEnvironmentsForLineno{gather}%
  \patchBothAmsMathEnvironmentsForLineno{multline}%
}

\definecolor{tabcolor}{rgb}{.410,.10,.11}
\usepackage[unicode=true, pdfusetitle,
 bookmarks=true, bookmarksnumbered=false, bookmarksopen=false,
 breaklinks=false, pdfborder={0 0 0}, pdfborderstyle={}, backref=false, colorlinks=true]
 {hyperref}
\hypersetup{citecolor=blue}

\usepackage{soul, txfonts}

\usepackage{stackengine}

\usepackage{diagbox}
\usepackage{fontspec}
\usepackage{ulem}

\soulregister\cite7
\soulregister\ref7
\soulregister\pageref7

\makeatletter

\begin{document}

\author{Jia-Nan Wu}
\affiliation{School of Physics, Beijing Institute of Technology, Beijing 100081, China.}
\author{Bingsuo Zou}
\affiliation{MOE $\&$ Guangxi Key Laboratory of Processing for Non-ferrous Metals and Featured Materials, School of Physical science and Technology, Guangxi University, Nanning 530004, China.}
\author{Guojun Jin}
\email{gjin@nju.edu.cn}
\affiliation{School of Physics Science and Technology, Kunming University, Kunming 650214, China.}
\affiliation{National Laboratory of Solid State Microstructures, Department of Physics, and Collaborative Innovation Center of Advanced Microstructures, Nanjing University, Nanjing 210093, China.}
\author{Yongyou Zhang}
\email{yyzhang@bit.edu.cn}
\affiliation{Beijing Key Laboratory of Nanophotonics $\&$ Ultrafine Optoelectronic Systems, School of Physics, Beijing Institute of Technology, Beijing 100081, China.}

\title{Robust Preparation of Bell States by Time Boundary Effect}

\date{\today} 

\begin{abstract}
Quantum entanglement is essential for modern quantum information processing. 
Entanglement gates convert initially non-entangled states into entangled ones by applying time-dependent parametric pulses. 
While Bell state preparation has been experimentally validated in various platforms, its stability and fidelity are constrained by environmental decoherence and parametric fluctuations.
Here, we propose a dynamically robust method for preparing Bell states by leveraging the time boundary effect and momentum-space projective measurements within Su-Schrieffer-Heeger systems. 
Employing the Lindblad master equation, we theoretically demonstrate that the prepared Bell states exhibit remarkable robustness against both environmental decoherence and parametric time fluctuations, achieving nearly perfect quantum fidelity, with the momentum conservation law governing this robust behavior. 
This framework based on the time boundary effect is applicable to both fermionic and bosonic excitations, offering a unified paradigm for generating Bell states in quantum computation and communication.
\end{abstract}


\maketitle

\section{Introduction}
Quantum entanglement \cite{RevModPhys.81.865}, a fundamental signature of quantum mechanics, serves as the cornerstone for quantum information processing \cite{PhysRevA.76.012335, Slussarenko2019}. 
This nonclassical phenomenon enables quantum superdense coding \cite{Liu2002}, entanglement-based teleportation \cite{pirandola2015advances}, entanglement swapping \cite{Liu2022}, quantum key distribution protocols \cite{sasaki2014practical}, and quantum secret processing \cite{QuantumSecure2017}. 
Among various entangled states, two-qubit maximally entangled Bell states \cite{gisin1998bell, Kim2001QuantumTeleportation, Distinguishability2001, ji2020fast, Zhao2022Dissipative} occupy a privileged position due to their mathematical elegance. 
They, as fundamental bases of quantum computation \cite{PhysRevA.76.012335} and communication \cite{gisin2007quantum, Complete2010, dong2011controlled}, achieve the maximal von Neumann entropy.
In experiments, Bell states have been demonstrated across diverse physical platforms, including ion hyperfine states \cite{Clark2021High}, electron spin states \cite{zou2022Bell-state}, photon polarization states \cite{Kim2018Informationally, wu2023experimental} and different frequency states \cite{PhysRevA.98.062327}.
However, two critical challenges persist in implementations: (i) environmental decoherence processes that degrade quantum fidelity through system-bath coupling \cite{Zhao2022Dissipative};
(ii) parametric sensitivity in Hamiltonian engineering.
These considerations motivate the development of robust Bell state generation methods capable of simultaneously suppressing environmental decoherence and compensating for parametric time fluctuations.

\begin{figure*}
  \centering
  \includegraphics[width=0.98\textwidth]{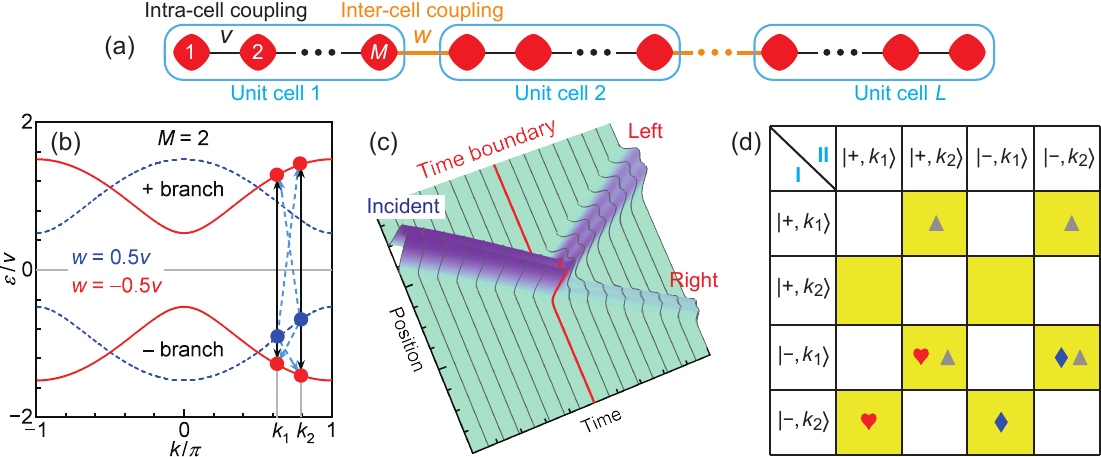}
  \caption{(a) 1D SSH model with $M$ sites per unit cell, intra-cell coupling $v$ (thin bonds), and inter-cell coupling $w$ (thick bonds). $L$ denotes the number of unit cells.
  (b) Band structures of the SSH model with $M=2$ as $w=0.5v$ (blue dashed lines, before time boundary) and $-0.5v$ (red solid lines, after time boundary). 
  (c) Schematics of the time boundary effect: the incident two-excitation state corresponding to the blue dots in (b) is scattered into the left- and right-moving waves, corresponding to the lower and upper two red dots in (b), respectively.
  (d) Sixteen two-excitation states $\hat{b}_{\alpha, k}^\dag \hat{b}_{\beta, k'}^\dag |0\rangle$ involved in the time boundary effect.}\label{Fig1}
\end{figure*}

Compared with conventional spatial modulations, such as quantum emitters coupled to waveguides \cite{Fan2009Theory} and curved topological boundaries \cite{Dong2023Reconfigurable}, and space-time modulations in classical optics \cite{akhmanov1969nonstationary, Kolner1994Space}, this work employs time-dependent Hamiltonians to dynamically reconfigure energy band structures without modifying the spatial period \cite{PhysRevE.64.026503, dong2024quantum}. 
Such an abrupt temporal discontinuity in a Hamiltonian is the so-called time boundary, whose effect has attracted broad interest these years \cite{zhou2020broadband, Wu2024Edge, Probing2025Xu}.
It is a deliberate design to control quantum wave dynamics, rather than the ``boundary time crystal'' \cite{Iemini2018Boundary} that appears in non-equilibrium phases of many-body systems due to spontaneous breaking of time-translation symmetry.
When a wave packet encounters a time boundary, the instantaneous eigenstates of the system change discretely, splitting the wave packet into multiple components, including time reflection and time refraction \cite{JTMendonca2002, Long2023time, Bar-Hillel2024Time}.
Time reflection exhibits group velocity reversal \cite{bacot2016time, moussa2023observation, agrawal2024temporal}, while time refraction maintains group velocity direction \cite{mendoncca2024time}.
The time boundary effect imposes strict momentum conservation, resulting in energy redistribution on band branches.
This energy-momentum decoupling has been experimentally validated in Su-Schrieffer-Heeger (SSH) momentum lattices \cite{dong2024quantum}, demonstrating high scattering efficiency. 
Other diverse platforms also revealed intriguing phenomena and applications of time boundary effects, including photonic time crystals \cite{Sharabi:22, Peng2022Topological} and Floquet metamaterials \cite{weitenberg2021tailoring}.
This transition from spatial to temporal control paradigms introduces unprecedented capabilities for addressing specific quantum challenges while maintaining robustness against environmental perturbations.
Breakthroughs in time-modulated control technologies motivate re-examination of Bell state preparation methodologies. 

In quantum information, entanglement gate operations use abrupt or gradual time-dependent parametric pulses to convert initially non-entangled states into entangled ones.
To demonstrate how the time boundary effect can be harnessed for Bell state preparation, we here consider a one-dimensional SSH lattice \cite{atala2013direct} with time-dependent inter-cell coupling
and report a robust preparation method for Bell states by the time boundary effect; see Fig.~\ref{Fig1}(a) with $M=2$.
Figure~\ref{Fig1}(b) shows the band structures before and after the time boundary, where the inter-cell coupling changes its value. Figures~\ref{Fig1}(c) and (d) illustrate the time boundary effect for the two-excitation states.
Such a framework not only finds applications to both fermionic and bosonic excitations, but also exhibits robustness against time-boundary variations, environmental decoherence, and parametric perturbations.
This robustness stems from the momentum conservation law rather than the topological properties inherent to the SSH model \cite{doi:10.34133/2019/8609875}.

This work is organized as follows.
In Sec.~\ref{sec2}, we introduce the model of the time boundary and derive the parametric conditions for the Bell state preparation.
In Sec.~\ref{sec3}, the robustness of the preparation is analyzed, including the gradual time boundary, environmental decoherence, and parametric fluctuations.
In Sec.~\ref{sec4}, we extend the SSH model to 
$M>2$ scenarios, which gives rise to multi-band time refraction/reflection and further enriches the available Bell state resources.
Finally, we summarize the main conclusions in Sec.~\ref{sec5}.

\section{Time boundary effect}\label{sec2}
\subsection{Model}

The considered SSH model in Fig.~\ref{Fig1}(a) consists of $L$ unit cells, each containing $M$ lattice sites, with intra-cell coupling $v$ (thin bonds) and inter-cell coupling $w$ (thick bonds).  
The Hamiltonian reads
\begin{align}\label{Hamiltonian}
\hat{\cal H} = \sum_{l=1}^L \sum_{m=1}^{M-1} v \hat{a}_{l,m}^\dag \hat{a}_{l,m+1} + \sum_{l=1}^{L-1} w(t) \hat{a}_{l,M}^\dag \hat{a}_{l+1,1} + \text{h.c.},  
\end{align}
where $\hat{a}_{l,m}^\dag$ ( $\hat{a}_{l,m}$ ) denotes the creation (annihilation) operator for a particle or excitation at the $m$-th site of the $l$-th unit cell. 
A time boundary could be realized by introducing time dependence into the inter-cell coupling, $w(t)$.
The intra-cell coupling $v$ is set as energy unit, with its inverse $v^{-1}$ as time unit.
For simplicity, the Planck constant is set to $\hbar\equiv1$ throughout this work.

Starting from the SSH model with $M = 2$, we analyze the time boundary effect.
The dispersion relations and eigenstates of $\hat{\mathcal{H}}$, respectively, read  
\begin{align}
\varepsilon_{\pm} &= \pm\sqrt{w^2 + v^2 + 2wv\cos k},  \\
|\pm, k\rangle &\equiv \left[g_{1,k}^{(\pm)}, \, g_{2,k}^{(\pm)} \right]^T = \frac{1}{\sqrt{2}} \left[ 1, \, \pm e^{i\phi_k} \right]^T,  
\end{align}
with lattice constant as length unit and phase factor $e^{i\phi_k} \equiv \frac{1 + \eta e^{ik}}{\left|1 + \eta e^{ik}\right|}$, where $\eta \equiv w/v$ represents the normalized inter-cell coupling.  
The band structure evolves with $w$; see Fig.~\ref{Fig1}(b) for $w = 0.5v$ (blue lines) and $w = -0.5v$ (red lines).  
An abrupt transition of the inter-cell coupling $w$ from $0.5v$ to $-0.5v$ at time $t_c$ induces a splitting of an initial wave into two scattered components.
This scattering phenomenon obeys momentum conservation.
In Fig.~\ref{Fig1}(b), black solid arrows depict the splitting processes of single-excitation waves at blue dots into the scattered ones at the upper and lower red dots.
For double-excitation waves, additional cross-splitting processes occur; see the blue dashed arrows.
For the red bands in Fig.~\ref{Fig1}(b), scattered waves on the upper branch propagate to the right, whereas those on the lower branch propagate to the left.  
Figure~\ref{Fig1}(c) exemplifies a two-excitation incident pulse split into right- and left-propagating waves due to the time boundary effect.  

\begin{figure*}
  \centering
  \includegraphics[width=0.98\textwidth]{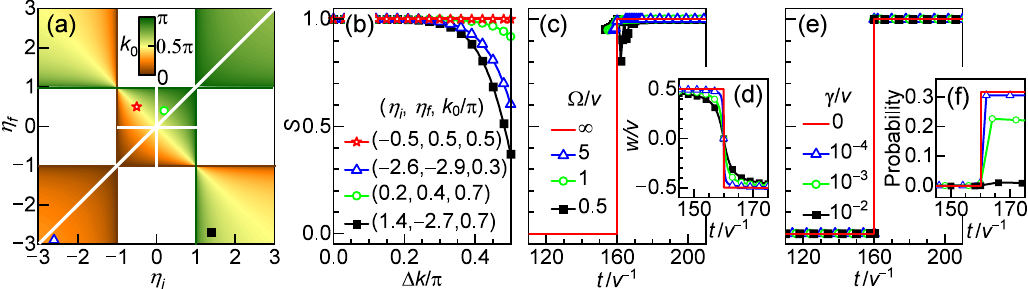}
\caption{(a) Parameter regime for achieving Bell states.  
  (b) Variations of the von Neumann entropy $S$ of $|f_{\rm M}\rangle$ with the wave vector difference $\Delta k$ of two excitations at the four points marked in (a).  
  (c) Time evolution of $S$ under (d) four types of time boundaries $w(t)$, illustrating the robustness of entanglement against gradual boundary change.
  (e) Time evolution of $S$ under four different decoherence rates $\gamma$, indicating the entanglement robustness against environmental decoherence, and (f) the corresponding Bell state probabilities. 
  Parameters: in (a, b), step time boundaries are applied with $\gamma=0$; in (c-f), $k_0 = 0.5\pi$, $\Delta k = 0.02\pi$, $k_w = 0.08\pi$, and $t_c = 160v^{-1}$; (c, d) assume no decoherence, while (e, f) account for decoherence and incorporate step time boundaries with $\eta_i=-\eta_f=0.5$; the fermionic SSH model ($M=2$) is used.}  
\label{Fig2}
\end{figure*}

\subsection{Parametric conditions for Bell states}\label{sec2B}
To investigate the parametric regime for preparing Bell states, we take the two-excitation plane wave  
\begin{align}
|i\rangle = \hat{a}_{-,k_1}^\dagger \hat{a}_{-,k_2}^\dagger |0\rangle  
\label{plwave}
\end{align}
as the incident state, see the two blue dots in Fig.~\ref{Fig1}(b), where $|0\rangle$ denotes the vacuum state with no particle across all sites. 
The operator $\hat{a}_{\alpha,k}^\dagger$ creates the eigenmode corresponding to branch $\alpha$ and wave vector $k$, prior to $t_c$.  
Here, the condition of $k_1 \neq k_2$ is assumed, since the case $k_1=k_2$ in $|i\rangle$ is forbidden. 
For fermions, the anticommutation relation leads to $|i\rangle=0$ as $k_1=k_2$.
For bosons, though the condition $k_1=k_2$ is acceptable, we will later show that it does not induce entanglement.
The state $|i\rangle$, which is marked by the two blue diamonds in Fig.~\ref{Fig1}(d),
evolves into the final state
\begin{align}\label{final state}
|f\rangle = \sum_{\alpha\beta, kk'} \psi_{ k,k'}^{(\alpha,\beta)} \hat{b}_{\alpha, k}^\dag \hat{b}_{\beta, k'}^\dag |0\rangle.
\end{align}
where $\hat{b}_{\alpha, k}^\dag$ denotes the eigenmode creation operator, subsequent to $t_c$.
Across time $t_c$, the normalized inter-cell coupling $\eta$ transitions from the initial value $\eta_i$ to the final value $\eta_f$.
Sixteen possible two-excitation momentum states emerge as $|\alpha, k\rangle|\beta, k'\rangle\ \left(\equiv\hat{b}_{\alpha, k}^\dag \hat{b}_{\beta, k'}^\dag |0\rangle\right)$ after time $t_c$, since $\alpha, \beta \in \{+,-\}$ and $k, k' \in \{k_1, k_2\}$, see Fig.~\ref{Fig1}(d). 
However, momentum conservation restricts them to eight states of the form $|\alpha, k_1\rangle|\beta, k_2\rangle$ and $|\alpha, k_2\rangle|\beta, k_1\rangle$ with $\alpha, \beta \in \{+,-\}$, highlighted by the yellow backgrounds. 
For bosonic and fermionic excitations, their coefficients satisfy $\psi_{k_1,k_2}^{(\alpha,\beta)} = +\psi_{k_2,k_1}^{(\beta, \alpha)}$ and $\psi_{k_1,k_2}^{(\alpha,\beta)} = -\psi_{k_2,k_1}^{(\beta, \alpha)}$, respectively.

For an indistinguishable two-excitation state, leveraging the quantum resources encoded in $|f\rangle$ presents a challenge. 
Here we employ projective measurements to characterize operational entanglement \cite{Lo2018Indistinguishability, Monkman2020Operational}.  
For $|f\rangle$, two approaches can be used to define projective measurement operators. 
The first utilizes momenta to distinguish excitations, designating those with $k_1$ and $k_2$ as particles I and II, respectively. The corresponding projection operator is  
$
\hat{Q}_{\rm B} = \sum_{\alpha, \beta\in\{+, -\}} |\alpha\rangle_{\rm I} |\beta\rangle_{\rm II} \, {}_{\rm II}\!\langle\beta| \, {}_{\rm I}\langle\alpha|,  
$
where $|\alpha\rangle_{\rm I} \equiv |\alpha, k_1\rangle$ and $|\beta\rangle_{\rm II} \equiv |\beta, k_2\rangle$. 
$\hat{Q}_{\rm B}$ measures the branch-correlation between particles I and II, encompassing four states highlighted by the gray triangles in Fig.~\ref{Fig1}(d). 
These states indicate that the measurement density operator
 ${\hat \rho}=\hat{Q}_{\rm B}|f\rangle\langle f|\hat{Q}_{\rm B}^\dag$ does not guarantee high entanglement; see Eq.~\eqref{reduce_den_mat}. 
 
The alternative approach distinguishes excitations by band branches, assigning those on the lower and upper branches as particles I and II, respectively. The associated projection operator reads
\begin{align}
\hat{Q}_{\rm M} = \sum_{k, k' \in \{k_1, k_2\}} |k\rangle_{\rm I} |k'\rangle_{\rm II} \, {}_{\rm II}\langle k'| \, {}_{\rm I}\langle k|,  
\end{align}
where $|k\rangle_{\rm I} \equiv |-, k\rangle$ and $|k'\rangle_{\rm II} \equiv |+, k'\rangle$. 
Momentum conservation restricts $\hat{Q}_{\rm M}$ to two states $|k_1\rangle_{\rm I} |k_2\rangle_{\rm II}$ and $|k_2\rangle_{\rm I} |k_1\rangle_{\rm II}$, marked by the two red hearts in Fig.~\ref{Fig1}(d).  
Owing to distinct group velocities of particles I and II, $\hat{Q}_{\rm M}$ exhibits at least two advantages:  
(i) spatial separability of particles I and II, see Fig.~\ref{Fig1}(c);
(ii) the measurement density operator  $\hat{\rho}_{\rm M} = \hat{Q}_{\rm M}|f\rangle\langle f|\hat{Q}_{\rm M}^\dagger$ can characterize a Bell state in momentum space.  
To determine the condition, we express the projection of $|f\rangle$ by $\hat{Q}_{\rm M}$ as  
\begin{align}\label{f12}  
|f_{\rm M}\rangle = \hat{Q}_{\rm M}|f\rangle = \psi_{k_1,k_2} |k_1\rangle_{\rm I} |k_2\rangle_{\rm II} + \psi_{k_2,k_1} |k_2\rangle_{\rm I} |k_1\rangle_{\rm II}.
\end{align}  
where $\psi_{k, k'}=2\psi_{k, k'}^{(-,+)}$. This expression shows that if \(k_1 = k_2\), the two projected bases $|k_1\rangle_{\rm I} |k_2\rangle_{\rm II}$ and $|k_2\rangle_{\rm I} |k_1\rangle_{\rm II}$ are identical.
Therefore, $|f_{\rm M}\rangle$ becomes a product state, implying no entanglement.
The condition $|\psi_{k_1,k_2}| = |\psi_{k_2,k_1}|$ with $k_1\neq k_2$ makes $|f_{\rm M}\rangle$ a Bell state.  
Setting $k_1 = k_0 - \Delta k/2$ and $k_2 = k_0 + \Delta k/2$, the resulting parametric equation  
\begin{align}\label{bellcondition}  
\cos k_0 = -\frac{\eta_i + \eta_f}{1 + \eta_i\eta_f} \quad (\eta_i \neq \eta_f;\ \eta_i, \eta_f \notin \{0,1\})  
\end{align}  
ensures, up to $O(\Delta k)$, $\psi_{k_1,k_2} = \pm\psi_{k_2,k_1}$ for bosonic and fermionic excitations, respectively.
The solution regime of Eq.~\eqref{bellcondition}, see Fig.~\ref{Fig2}(a), satisfies either $|\eta_i| < 1$ and $|\eta_f| < 1$, or $|\eta_i| > 1$ and $|\eta_f| > 1$. 
In particular, when 
\begin{align}\label{bc}  
\eta_i + \eta_f =0, \quad k_0 = \frac{\pi}{2}, 
\end{align} 
the relation $\psi_{k_1,k_2} \equiv \pm\psi_{k_2,k_1}$ holds for any $\Delta k$ and thus $|f_{\rm M}\rangle$ becomes an ideal Bell state.
The detailed derivation for Eqs.~\eqref{bellcondition} and \eqref{bc} is presented in Appendix~\ref{AppA}.

The four points in Fig.~\ref{Fig2}(a) are selected to plot the von Neumann entropy of $|f_{\rm M}\rangle$,  
\begin{align}
S = -|\psi_{k_1,k_2}|^2\log_2|\psi_{k_1,k_2}|^2 - |\psi_{k_2,k_1}|^2\log_2|\psi_{k_2,k_1}|^2,  
\end{align}
as a function of $\Delta k$ in Fig.~\ref{Fig2}(b). When $\eta_i = -\eta_f$ and $k_0 = \pi/2$, the entropy strictly equals 1; see the star point in Fig.~\ref{Fig2}(a) and the red star-dotted line in Fig.~\ref{Fig2}(b). 
For the other three points, the entropies remain close to the maximum entanglement limit when $\Delta k < 0.2\pi$. 
Therefore, the parametric regime is sufficiently broad to facilitate Bell state preparation. 

In addition to tuning $\eta$, tuning onsite energies can also give rise to Bell states with high entanglement, as demonstrated in Appendix~\ref{AppB}.

\section{Robustness}\label{sec3}
This section discusses the robustness of the preparation of Bell states by the time boundary effect.
Three perturbations are considered, i.e., gradual time boundary (Sec.~\ref{sec3A}), environmental decoherence (Sec.~\ref{sec3B}), and parametric fluctuations (Sec.~\ref{sec3C}).
Here, robustness means that the prepared Bell states maintain maximum entanglement (the von Neumann entropy $S=1$) and high quantum fidelity under these imperfect influences. 

\subsection{Gradual time boundary}\label{sec3A}
The discussions in Figs.~\ref{Fig2}(a) and \ref{Fig2}(b) are based on the ideal step time boundary, i.e., $\eta$ encounters an abrupt change at $t_c$, being far from the adiabatic limit.
To explore the impact of the gradual time boundary, the Gaussian wavefunction of
\begin{align}\label{gstate}  
|i_G\rangle = \sum_{l_1,m_1,l_2,m_2} \psi_{l_1, m_1}\psi_{l_2, m_2} \hat{a}_{l_1,m_1}^{\dag}\hat{a}_{l_2,m_2}^{\dag}|0\rangle,  
\end{align}
is taken as the initial state in this subsection. Its amplitude, $\psi_{l_n, m_n}$ with $n\in\{1,2\}$, has the form:
\begin{align}
\psi_{l_n,m_n} \propto \sum_{k} e^{-\frac{(k - k_{n})^2}{k_w^2}} g_{m_n,k}^{(-)} e^{ik(l_n - x_{0,n})},  
\end{align}
characterized by a wave vector width $k_w$, two wave vector centers $k_n$, and two initial position centers $x_{0,n}$.
After the gradual time boundary, we can calculate the von Neumann entropy $S$ of the projective state $|f_{\rm M}\rangle$. 
The gradual time boundary is mathematically described by
\begin{align}\label{gtb}
w(t)/v = -\frac{1}{\pi} \arctan\left[\Omega(t - t_c)\right],  
\end{align}
where, as $\Omega$ increases from 0 to $\infty$, the time boundary smoothly transforms into a step one; see Fig.~\ref{Fig2}(d).
Since the regime for the time boundary effect is far from the adiabatic limit, the value of $\Omega$ is studied from $0.5v$ to $\infty$, as shown in Fig.~\ref{Fig2}(d).
In Appendix~\ref{AppC}, we prove that $|f_{\rm M}\rangle$ in Eq.~\eqref{f12} is always a maximum-entangled pure state under Eq.~\eqref{bc}, regardless of the value of $\Omega$ in Eq.~\eqref{gtb}.
Figure~\ref{Fig2}(c) confirms this numerically, that is, the von Neumann entropy approaches 1 after time $t_c=160v^{-1}$.
This property is of crucial significance for experimental realizations.

\begin{figure}
  \centering
  \includegraphics[width=0.48\textwidth]{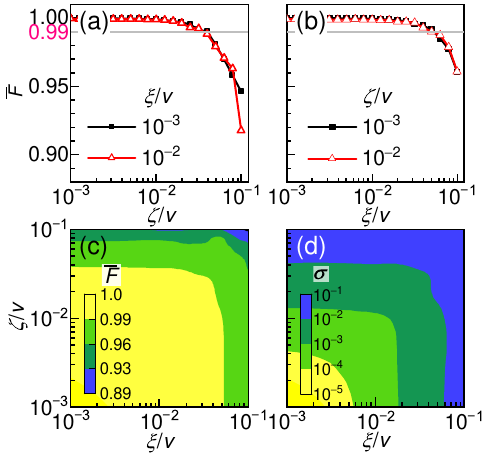}
  \caption{Statistical averages of fidelity $\bar{F}$ for $|f_{\rm M}\rangle_{\rm Rand}$ at $t = 320v^{-1}$, varying with (a) $\zeta$ under different $\xi$ and (b) $\xi$ under different $\zeta$.
  Contour maps of (c) $\bar{F}$ and (d) standard deviation $\sigma$ in the $\zeta\text{-}\xi$ space for $|f_{\rm M}\rangle_{\rm Rand}$ at $t = 320v^{-1}$.
Parameters: $\eta_i = -\eta_f = 0.5$, $k_0 = 0.5\pi$, $\Delta k = 0.02\pi$, $k_w = 0.08\pi$, and $t_c = 160v^{-1}$; the fermionic SSH model ($M=2$) is used.}\label{Fig3}
\end{figure}
\subsection{Environmental decoherence}\label{sec3B}

Environmental decoherence is another challenge to the realization of reliable quantum processing.  
Useful techniques to mitigate decoherence-induced effects include quantum information masking \cite{Liu2021Photonic}, quantum error correction \cite{Concatenating2024Dash}, dynamical decoupling \cite{lidar2014review}, and adopting decoherence-free subspaces \cite{lidar2014review, Concatenating2024Dash, Deterministic2025Rubies}.  
To measure the influence of environmental decoherence on $|f_{\rm M}\rangle$ in Eq.~(\ref{f12}), we consider the Markovian loss, described by the Lindblad master equation,
\begin{align}
\frac{d\hat{\rho}}{dt} = -i\left[\hat{\cal H},\ \hat{\rho}\right] + \gamma\sum_{l,m} \left( \hat{a}_{l,m}\hat{\rho}\hat{a}_{l,m}^{\dag} - \frac{1}{2}\left\{ \hat{\rho}, \hat{a}_{l,m}^{\dag}\hat{a}_{l,m} \right\} \right),\label{master equation}  
\end{align}
where $\gamma$ quantifies the loss, i.e., the decoherence rate \cite{Simon2022Lindblad}.

Using the incident pulse $\hat{\rho}_0 = |i_G\rangle\langle i_G|$ defined by Eq.~\eqref{gstate},
we can solve this master equation by the fourth-order Runge-Kutta method. 
The numerical results show that the measurement density operator, ${\hat \rho}_{\rm M} = \hat{Q}_{\rm M} {\hat \rho} \hat{Q}_{\rm M}^\dagger$ always describes a pure state, even as ${\hat \rho}$ becomes mixed due to decoherence.
The corresponding von Neumann entropy of $\hat\rho_{\rm M}$ is illustrated in Fig.~\ref{Fig2}(e) under a step time boundary at $t_c=160v^{-1}$.
Clearly, ${\hat \rho}_{\rm M}$ holds the maximum entanglement and is irrespective of $\gamma$. However, $\gamma$ substantially affects the success probability of the preparation, see Fig.~\ref{Fig2}(f), which is measured by $\rm{Tr}(\hat{\rho}_M)$. 
As $\gamma$ increases from 0 to $0.001v$ and $0.01v$, the success probability drops from 31.6\% to ${\sim}20\%$ and ${\sim}1\%$.
This is physically intuitive, since the non-zero $\gamma$ reduces the number of excitations within the system. 
While decoherence decreases the success probability, it exerts a negligible influence on the entanglement of the prepared states.
These prepared states hold maximum entanglement and are therefore robust against decoherence. 

In fact, this entanglement robustness against decoherence can be proved analytically by analyzing the action of the Lindblad dissipation on all involved Fock states, see Appendix~\ref{AppD} for the proof.

\begin{figure*}
  \centering
  \includegraphics[width=0.9\textwidth]{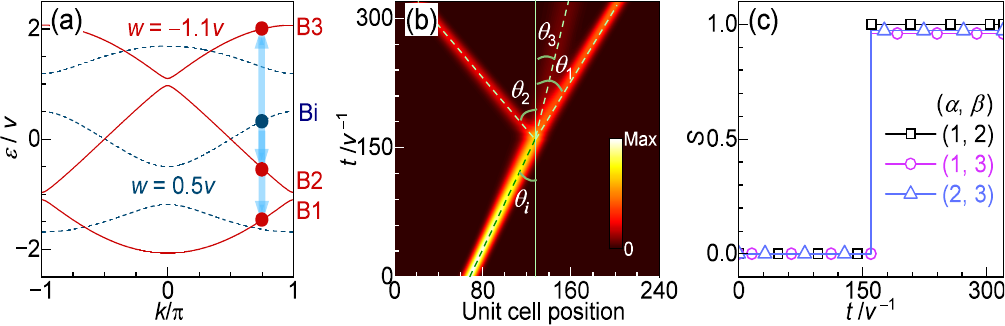}
  \caption{(a) Band structures of the three-band system ($M=3$) with $w=0.5v$ (before time boundary) and $w=-1.1v$ (after time boundary).
(b) Schematic of the three-band time refraction and reflection.
(c) Time evolution of $S$ for the projected states that two particles distribute on different bands.
Parameters: $k_{0}=0.75\pi$, $\Delta k=0.02\pi$ $k_w=0.08\pi$, $x_0=69$, $\eta_i=0.5$, $\eta_f=-1.1$, $t_c=160v^{-1}$; the fermionic SSH model ($M=3$) is used.}\label{Fig4}
\end{figure*}

\subsection{Parametric fluctuation}\label{sec3C}
Experimental systems inherently suffer from parametric time fluctuations that may affect eigenstates and band structures during time evolution. 
Thus, it is also critical to evaluate how such parametric fluctuations impact the prepared Bell states. This subsection focuses on two types of parametric fluctuations, that is, hoppings and onsite energies. 

Without loss of generality, the perturbation amplitudes for the two hoppings of $v$ and $w$ are assumed identical, denoted by $\zeta$, and that for the onsite energies is denoted by $\xi$. 
Note that the onsite energies are set to zero in Eq.~\eqref{Hamiltonian}. 
Uncorrelated random fluctuations are modeled as $v \rightarrow v + \zeta \cdot \text{Rand}(t)$, $w \rightarrow w + \zeta \cdot \text{Rand}(t)$, and $\text{onsite energies} \rightarrow \xi \cdot \text{Rand}(t)$, where $\text{Rand}(t)$ is a random number in $(-1, 1)$. 
Using the initial Gaussian state of Eq.~(\ref{gstate}) and the projection operator $\hat{Q}_{\text{M}}$, we characterize the final state $|f_{\rm M}\rangle_{\text{Rand}}$ under parametric perturbations by the quantum fidelity, $F = \left|\langle f_{\rm M} | f_{\rm M}\rangle_{\rm Rand}\right|^2$,  
where $|f_{\rm M}\rangle$ corresponds to the unperturbed state ($\zeta = \xi = 0$).  
Figures~\ref{Fig3}(a) and \ref{Fig3}(b) illustrate the variations of the statistical average of quantum fidelity $\bar{F}$ with respect to $\zeta$ and $\xi$, respectively.  
For statistical average, 100 disordered time evolution processes are used. 
As expected, large parametric fluctuations substantially degrade the fidelity. 
However, when $\zeta$ and $\xi$ are below $\sim$0.04$v$, $\bar{F}$ remains above 0.99, indicating that $|f_{\rm M}\rangle_{\rm Rand}$ closely approximates the ideal Bell state.  
Figure~\ref{Fig3}(c) presents a contour map of $\bar{F}$ in the $\zeta$-$\xi$ space.
The yellow region ($\bar{F} > 0.99$) confirms the robustness of $|f_{\rm M}\rangle_{\rm Rand}$ against parametric time perturbations.  
Correspondingly, the standard deviation of $F$, $\sigma$, shown in Fig.~\ref{Fig3}(d), is less than 0.01, further validating the stability of the preparation scheme.
If the fidelity threshold is relaxed to $\bar{F} > 0.9$, the parametric fluctuations up to $\sim$0.1$v$ are acceptable.
Accordingly, the preparation scheme based on the time boundary effect does not rely on fine-tuned parameters.

\section{Multi-band time boundary}\label{sec4}
When generalizing the SSH model with $M=2$ studied above to a multi-band system by adding more sites per unit cell (i.e., $M>2$), we can achieve multi-band time boundaries and thus may enrich entanglement resources from more pairs of energy bands.
As an instance, the band structures with $M = 3$ for $w = 0.5v$ (dashed lines) and $w = -1.1v$ (solid lines) are shown in Fig.~\ref{Fig4}(a). 

Similar to the $M = 2$ case, when $w$ undergoes an abrupt transition from $0.5v$ to $-1.1v$, a step time boundary is formed. 
In Fig.~\ref{Fig4}(a), the initial state $|i\rangle$ corresponds to the dark-blue dot on the band `Bi'. 
Due to the time boundary, this state is scattered into three energy bands of B1, B2, and B3, see Fig.~\ref{Fig4}(a). Momentum conservation leads to the final states primarily accumulating around the three red dots. 
The group velocities of the three bands near the red dots exhibit different values and thus give rise to different time refraction or reflection.  
As illustrated in Fig.~\ref{Fig4}(b), an initial Gaussian single-excitation state is split into three branches. 
Here, $\theta_i$ denotes the incident angle relative to the time boundary. 
Additionally, $\theta_1$, $\theta_2$, and $\theta_3$ represent the angles corresponding to bands B1, B2, and B3, respectively. 
Similar to Fig.~\ref{Fig1}(c), these angles are determined by the group velocities of the eigenstates in each band, see Appendix~\ref{angles}.

Bell states in momentum space can be prepared on any pair of energy bands. 
Taking the initial two-excitation Gaussian state like Eq.~(\ref{gstate}) on the `Bi' band as an example, the parameters are set as $k_0 = 0.75\pi$, $\Delta k = 0.02\pi$, $\eta_i = 0.5$, and $\eta_f = -1.1$. 
The projective operator $\hat{Q}_{\text{M}}$ is extended to $\hat{Q}_{\text{M}}^{(\alpha, \beta)}$ to accommodate three energy bands (B1, B2, B3), 
\begin{align}
\hat{Q}_{\rm M}^{(\alpha, \beta)} = \sum_{k, k'} |\alpha, k\rangle_{\rm I} |\beta, k'\rangle_{\rm II} \, {}_{\rm II}\langle \beta, k'| \, {}_{\rm I}\langle \alpha, k|,  
\end{align}
where $\alpha, \beta \in \{1, 2, 3\}$.  
After the time boundary, the von Neumann entropies $S$ for the projected states obtained by $\hat{Q}_{\text{M}}^{(1,2)}$, $\hat{Q}_{\text{M}}^{(1,3)}$, and $\hat{Q}_{\text{M}}^{(2,3)}$ are shown in Fig.~\ref{Fig4}(c).
All of them exhibit high entanglement. 
In Appendix \ref{AppE2}, the variations of $S$ with $\eta_i$ and $\eta_f$ are plotted for $\hat{Q}_{\text{M}}^{(1,2)}$, $\hat{Q}_{\text{M}}^{(1,3)}$, and $\hat{Q}_{\text{M}}^{(2,3)}$. 
Since the total number of $\{\alpha, \beta\}$ combinations increases with the band number $M$ as $M(M-1)/2$, the SSH model with $M>2$ significantly enriches available entangled states.

\section{Conclusion and discussion}\label{sec5}

In this work, we introduce time boundaries into the SSH model and utilize projective measurements to prepare Bell states in momentum space, whose parametric conditions are analytically derived.
The entanglement of the prepared Bell states is robust against system parameter oscillations and environmental decoherence.
This robustness is rooted in the momentum conservation law.
The prepared Bell states maintain high fidelity even under significant parametric time fluctuations.
Specifically, the fidelity of the prepared Bell states exceeds $99\%$ even when the stochastic disturbances of onsite energies or inter-site coupling strengths reach $4\%$.
Our study establishes a dynamical framework for the preparation of Bell states, applicable to both fermionic and bosonic excitations, thereby advancing potential applications in quantum technologies.

Noticeably, topologically protected edge states in two-dimensional (2D) lattices, characterized by robustness against local perturbations \cite{Dong2023Reconfigurable}, have been extensively investigated. 
Simultaneously, the interplay between band topology and time boundary effect has been established \cite{Wu2024Edge}. These foundational studies indicate that the dynamical framework for preparing robust Bell states could also achieve topological robustness against local stochastic disturbances when implemented in topological systems. 

\section*{Acknowledgement}
This work was supported by the National Natural Science 
Foundation of China (Grant Nos. 92570115 and 12074156).


\appendix
\section{Parametric conditions for Bell states}\label{AppA}
This section derives the parametric conditions of Bell states presented in Eq.~\eqref{bellcondition}. 
We start from the initial state given by Eq.~\eqref{plwave}. 
When encountering the time boundary, the eigenstates of the system transform from 
\begin{align}\label{eig vec i}
\left(\begin{matrix}g^{(-)}_{1, k}\\g^{(-)}_{2, k}\end{matrix}\right)=\left(\begin{matrix}-\frac{1+\eta_i e^{-ik}}{\sqrt{1+\eta_i^2+2\eta_i\cos k}}\\1\end{matrix}\right),\ \ 
\left(\begin{matrix}g^{(+)}_{1, k}\\g^{(+)}_{2, k}\end{matrix}\right)=\left(\begin{matrix}\frac{1+\eta_i e^{-ik}}{\sqrt{1+\eta_i^2+2\eta_i\cos k}}\\1\end{matrix}\right),
\end{align}
 into
 \begin{align}\label{eig vec f}
 \left(\begin{matrix}h^{(-)}_{1, k}\\h^{(-)}_{2, k}\end{matrix}\right)= \left(\begin{matrix}-\frac{1+\eta_f e^{-ik}}{\sqrt{1+\eta_f^2+2\eta_f\cos k}}\\1\end{matrix}\right),\ \ 
\left(\begin{matrix}h^{(+)}_{1, k}\\h^{(+)}_{2, k}\end{matrix}\right)=\left(\begin{matrix}\frac{1+\eta_f e^{-ik}}{\sqrt{1+\eta_f^2+2\eta_f\cos k}}\\1\end{matrix}\right).
 \end{align}
After time $t_c$, the initial state in Eq.~\eqref{plwave} transforms into that in Eq.~\eqref{final state}, where the coefficient $\psi_{ k,k'}^{(\alpha,\beta)}$ can be found by $\psi_{ k,k'}^{(\alpha,\beta)}=\left\langle0\left| \hat{b}_{\beta, k'} \hat{b}_{\alpha, k}\right|i\right\rangle$.
Using $\hat{b}_{\alpha, k}^{\dag}=\frac{1}{\sqrt{L}}
\sum_{l,m}\hat{a}_{l,m}^{\dag}h_{m, k}^{(\alpha)}e^{ikl}$, 
we obtain
\begin{align} \label{phi_ak1bk2}
\psi_{k, k'}^{(\alpha,\beta)}=\pm\psi_{k', k}^{(\beta,\alpha)}=&\sum_{m_1, m_2}\left[h_{m_2, k'}^{(\beta)\ast}
h_{m_1, k}^{(\alpha)\ast}\delta_{k_1k}\delta_{k_2k'}\right.\nonumber \\ 
&\left.\pm 
h_{m_1, k'}^{(\beta)\ast}
h_{m_2, k}^{(\alpha)\ast}\delta_{k_1k'}\delta_{k_2k}\right]
g_{m_1, k_1}^{(-)}g_{m_2, k_2}^{(-)},
\end{align}
where $+$ ($-$) is for bosonic (fermionic) excitations obeying the operator commutation (anticommutation) relation.

For $\hat{Q}_{\rm{B}}$, the post-measurement state reads
\begin{align}
|f_{\rm B}\rangle = \hat{Q}_{\rm B}|f\rangle = &\phi_{-,-} |-\rangle_{\rm I} |-\rangle_{\rm II}
+\phi_{-,+} |-\rangle_{\rm I} |+\rangle_{\rm II}\\\nonumber
&+\phi_{+,-} |+\rangle_{\rm I} |-\rangle_{\rm II}
+\phi_{+,+} |+\rangle_{\rm I} |+\rangle_{\rm II},  
\end{align}
where $\phi_{\alpha, \beta}=2\psi_{k_1, k_2}^{(\alpha,\beta)}$.
The post-measurement density matrix has the form
\begin{align}
\hat{\rho}_\textrm{B}&=|f_{\rm B}\rangle\langle f_{\rm B}|\\\nonumber
&=
\left(
  \begin{array}{cccc}
    |\phi_{-,-}|^2 & \phi_{-,-}\phi_{-,+}^* & \phi_{-,-}\phi_{+,-}^* & \phi_{-,-}\phi_{+,+}^* \\
     \phi_{-,+}\phi_{-,-}^* & |\phi_{-,+}|^2 & \phi_{-,+}\phi_{+,-}^* & \phi_{-,+}\phi_{+,+}^* \\\phi_{+,-}\phi_{-,-}^* & \phi_{+,-}\phi_{-,+}^* & |\phi_{+,-}|^2 & \phi_{+,-}\phi_{+,+}^* \\
    \phi_{+,+}\phi_{-,-}^* & \phi_{+,+}\phi_{-,+}^* & \phi_{+,+}\phi_{+,-}^* & |\phi_{+,+}|^2 \\
  \end{array}
\right),
\end{align}
and thus the reduced density matrix for particle I is
\begin{align} \label{reduce_den_mat}
\tilde{\rho}_\textrm{B}&=\sum_{\beta}{}_{\textrm{II}}\langle\beta|
\hat{\rho}_\textrm{B}
|\beta\rangle_{\textrm{II}}\\\nonumber
&=\left(
  \begin{array}{cc}
    |\phi_{-,-}|^2+|\phi_{-,+}|^2 & \phi_{-,-}\phi_{+,-}^*+\phi_{-,+}\phi_{+,+}^* \\
    \phi_{+,-}\phi_{-,-}^*+\phi_{+,+}\phi_{-,+}^* & |\phi_{+,-}|^2+|\phi_{+,+}|^2 \\
  \end{array}
\right).
\end{align}
Substituting Eqs.~(\ref{eig vec i}-\ref{phi_ak1bk2}) into (\ref{reduce_den_mat}) results in $S=-\textrm{Tr}(\tilde{\rho}_\textrm{B}
\log_2\tilde{\rho}_\textrm{B})\equiv0$, indicating that the projective measurement $\hat{Q}_{\rm{B}}$ does not induce operational entanglement.

For $\hat{Q}_{\rm{M}}$, the corresponding post-measurement state reads
\begin{align}  
|f_{\rm M}\rangle = \hat{Q}_{\rm M}|f\rangle = \psi_{k_1,k_2} |k_1\rangle_{\rm I} |k_2\rangle_{\rm II} + \psi_{k_2,k_1} |k_2\rangle_{\rm I} |k_1\rangle_{\rm II}, 
\end{align}  
i.e., Eq.~\eqref{f12} in the Sec.~\ref{sec2B} with $\psi_{k, k'}=2\psi_{k, k'}^{(-,+)}$. 
Obviously, $|f_{\rm M}\rangle$ demonstrates momentum entanglement.
To let $|f_{\rm M}\rangle$ be a Bell state, the condition $|\psi_{k_1k_2}|=|\psi_{k_2k_1}|$ is required.
To find the solution of $|\psi_{k_1,k_2}|=|\psi_{k_2,k_1}|$, we perform Taylor expansions for $\psi_{k_1,k_2}$, $\psi_{k_2,k_1}$ up to the first order of $\Delta k=k_2-k_1$, that is,
\begin{align}
\psi_{k_1,k_2}&\approx \left[2-2e^{2i\Xi(k_0)}\right]
+8ie^{i\Xi(k_0)}
{\cal Z}(k_0, \eta_i, \eta_f)\cdot\Delta k,\\
\psi_{k_2,k_1}&\approx (-1)^{q} \cdot \left[2-2e^{2i\Xi(k_0)}\right]\nonumber\\
&\quad+(-1)^{1-q} \cdot 8ie^{i\Xi(k_0)}
{\cal Z}(k_0, \eta_i, \eta_f)\cdot\Delta k,
\end{align}
where
\begin{align}
k_0&={1\over2}(k_1+k_2),\\
\Xi(k_0)&\equiv\arctan {\eta_f\sin k_0\over 1+\eta_f\cos k_0 } - \arctan {\eta_i\sin k_0\over 1+\eta_i\cos k_0 },\\
{\cal Z}(k_0, \eta_i, \eta_f)&\equiv\frac{\eta_f\cos k_0+\eta_f^2}{1+\eta_f^2+2\eta_f\cos k_0}-\frac{\eta_i\cos k_0+\eta_i^2}{1+\eta_i^2+2\eta_i\cos k_0}.
\end{align}
Here, we use $q=0$ for bosonic excitations, while $q=1$ for fermionic excitations. 

Setting ${\cal Z}(k_0, \eta_i, \eta_f)=0$, we have $\psi_{k_1,k_2}=+\psi_{k_2,k_1}$ for bosonic excitations and $\psi_{k_1,k_2}=-\psi_{k_2,k_1}$ for fermionic ones, up to \( O(\Delta k) \).
The equation of ${\cal Z}(k_0, \eta_i, \eta_f)=0$ leads to the parametric equation  
\begin{align}\label{bellconditiona}  
\cos k_0 = -\frac{\eta_i + \eta_f}{1 + \eta_i\eta_f} \quad (\eta_i \neq \eta_f;\ \eta_i, \eta_f \notin \{0,1\}),  
\end{align}
i.e., Eq.~\eqref{bellcondition} in Sec.~\ref{sec2B}.
The solution space is plotted in Fig.~\ref{Fig2}(a).
A strict solution can be found, that is, when $\eta_i=-\eta_f$ and $k_0={\pi\over2}$, one can strictly obtain $\psi_{k_1, k_2}^{(-,+)}=\pm\psi_{k_2, k_1}^{(-,+)}$ from Eqs.~(\ref{eig vec i}-\ref{phi_ak1bk2}), regardless of the value of $\Delta k$.
Namely, the condition of $\eta_i=-\eta_f$ and $k_0=\frac{\pi}{2}$ ensures that $|f_{\rm M}\rangle$ is strictly a Bell state.

\begin{figure}
  \centering
  \includegraphics[width=0.49\textwidth]{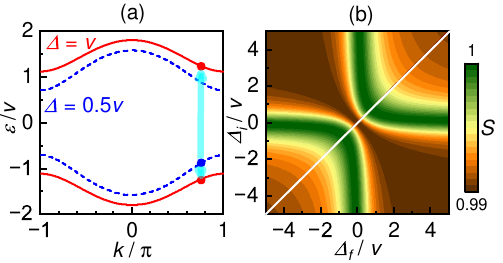}
  \caption{(a) Band structures of the SSH model ($M=2$) with $\Delta=0.5v$ (blue dashed lines) and $\Delta=v$ (red solid lines).
  (b) Variation of the von Neumann entropy for $|f_{\rm M}\rangle$ with $\Delta_i$ and $\Delta_f$. 
  Parameters: $k_0=0.75\pi$, $\Delta k=0.02\pi$, and $w=0.5v$.}\label{Fig5}
\end{figure}
  
\section{Time boundary by tuning onsite energies}\label{AppB}

In the main text, the time boundary is achieved by temporally tuning the inter-cell coupling strength $w$. 
This section shows an alternative way of tuning the onsite energies \cite{atala2013direct}. 
For this purpose, we here consider the SSH model with $M=2$ and write the Hamiltonian as
\begin{align}\label{Hamiltonian_d}
\hat{\cal H}=&\sum_{l}\left[v\hat{a}_{l,1}^{\dag}\hat{a}_{l,2}
+w\hat{a}_{l,2}^{\dag}\hat{a}_{l+1,1}+h.c.\right]\nonumber\\
&+\sum_{l}\Delta(t)\left[\hat{a}_{l,1}^{\dag}\hat{a}_{l,1}
-\hat{a}_{l,2}^{\dag}\hat{a}_{l,2}\right],
\end{align} 
where the onsite energy $\Delta(t)$ is time-dependent while $v$ and $w$ are constants.
The dispersion relations are $\varepsilon_\pm(k, \Delta)=\pm\sqrt{v^2+w^2+2vw\cos k+\Delta^2(t)}$.
When the time boundary is encountered, $\Delta(t)$ transitions from the initial value $\Delta_i$ to the final one $\Delta_f$.
Like Eqs.~\eqref{eig vec i} and \eqref{eig vec f}, when the system encounters the time boundary, the eigenstates change from 
\begin{align}\label{eig vec i2}
\left(\begin{matrix}g^{(-)}_{1,k}\\g^{(-)}_{2,k}\end{matrix}\right)=\left(\begin{matrix}
\frac{v+we^{-ik}}{\varepsilon_{-}(k, \Delta_i)-\Delta_i}\\1\end{matrix}\right),\quad
\left(\begin{matrix}g^{(+)}_{1,k}\\g^{(+)}_{2,k}\end{matrix}\right)=\left(\begin{matrix}
\frac{v+we^{-ik}}{\varepsilon_{+}(k, \Delta_i)-\Delta_i}\\1\end{matrix}\right)
\end{align}
to
 \begin{align}\label{eig vec f2}
\left(\begin{matrix}h^{(-)}_{1,k}\\h^{(-)}_{2,k}\end{matrix}\right)=\left(\begin{matrix}
 \frac{v+we^{-ik}}{\varepsilon_{-}(k, \Delta_f)-\Delta_f}\\1\end{matrix}\right),\quad
\left(\begin{matrix}h^{(+)}_{1,k}\\h^{(+)}_{2,k}\end{matrix}\right)=\left(\begin{matrix}
 \frac{v+we^{-ik}}{\varepsilon_{+}(k, \Delta_f)-\Delta_f}\\1\end{matrix}\right).
 \end{align}

Figure~\ref{Fig5}(a) shows the band transition as $\Delta$ changes from $\Delta_i=0.5v$ to $\Delta_f=v$.
Figure~\ref{Fig5}(b) presents the von Neumann entropies of the final states in the $\Delta_i$-$\Delta_f$ space with $k_1=0.74\pi$, $k_2=0.76\pi$, where $\Delta_i\neq\Delta_f$.
It clearly demonstrates that Bell states can also be effectively prepared through the time boundary effect by tuning onsite energies, since the von Neumann entropy approaches $1$.

\section{Proof of entanglement robustness against the gradual time boundary}\label{AppC}
In Sec.~\ref{sec3A}, we have numerically confirmed that the gradual time boundaries do not disrupt the prepared Bell states. 
This section confirms this phenomenon in theory by comparing two types of time processes. 
The first type involves $\eta$ sharply transforming from $\eta_i$ to $\eta_f$ at time $t_c$,
denoted as
\begin{align}\label{ft}
|i\rangle\xrightarrow[{{\rm at}\ t_c}]{\eta_i\rightarrow\eta_f}|f\rangle.
\end{align}
Like Eqs.~\eqref{plwave} and \eqref{final state}, the creation operators before and after $t_c$ are denoted as $\hat{a}_{\alpha, k}^\dag$ and $\hat{b}_{\alpha, k}^\dag$, respectively.

For the second type, $\eta$ firstly transforms from $\eta_i$ to $\eta_p$ at time $t_p$ and then to $\eta_f$ at time $t_c$, i.e.,
\begin{align}\label{st}
|i\rangle\xrightarrow[{{\rm at\ }t_p}]{\eta_i\rightarrow\eta_p}|p\rangle
\xrightarrow[{{\rm at\ } t_c}]{\eta_p\rightarrow\eta_f}|f\rangle.
\end{align}
In this case, we use $\hat{a}_{\alpha, k}^\dag$, $\hat{c}_{\alpha, k}^\dag$, and $\hat{b}_{\alpha, k}^\dag$ to denote the creation operators of the eigenmodes for $t$ in the intervals $(-\infty, t_p)$,  $(t_p, t_c)$, and $(t_c, +\infty)$, respectively. 

\subsection{First type}

For the time process in Eq.~\eqref{ft}, the $|i\rangle$ in Eq.~\eqref{plwave} is directly transformed into $  
|f\rangle = \sum_{\alpha\beta, kk'} \psi_{ k,k'}^{(\alpha,\beta)} \hat{b}_{\alpha, k}^\dag \hat{b}_{\beta, k'}^\dag |0\rangle$ at time $t_c$ with
\begin{align}\label{bc1}
\psi_{ k,k'}^{(\alpha,\beta)}=e^{-i\varepsilon_{i}t_c}
\left\langle0\left|\hat{b}_{\beta, k'}\hat{b}_{\alpha, k}\right|i\right\rangle
\end{align} 
where $\varepsilon_{i}$ is the energy of $|i\rangle$. 
According to Eq.~\eqref{f12}, the condition $\psi_{k_1,k_2}^{(-,+)}=\pm\psi_{k_2,k_1}^{(-,+)}$ leads to $|f\rangle$ being a Bell state.
\\

\subsection{Second type}

For the time process in Eq.~\eqref{st}, the initial state $|i\rangle$ is transformed into an intermediate state with the form $|p\rangle=\sum_{\mu\nu, \tilde{k}\tilde{k}'} \phi_{ \tilde{k},\tilde{k}'}^{(\mu,\nu)} \hat{c}_{\mu, \tilde{k}}^\dag \hat{c}_{\nu, \tilde{k}'}^\dag |0\rangle$ firstly and then transformed into $|f\rangle= \sum_{\alpha\beta, kk'} \psi_{ k,k'}^{(\alpha,\beta)} \hat{b}_{\alpha, k}^\dag \hat{b}_{\beta, k'}^\dag |0\rangle$.
The coefficient $\phi_{ \tilde{k},\tilde{k}'}^{(\mu,\nu)}$ has the form
\begin{align}\label{trans1}
\phi_{ \tilde{k},\tilde{k}'}^{(\mu,\nu)}=e^{-i\varepsilon_{i}t_p}
\left\langle0\left|\hat{c}_{\nu, \tilde{k}'}\hat{c}_{\mu, \tilde{k}}\right|i\right\rangle,
\end{align}
and its connection with $\psi_{ k,k'}^{(\alpha,\beta)}$ in $|f\rangle$ reads
\begin{align}\label{trans2}
\psi_{ k,k'}^{(\alpha,\beta)}=
\sum_{\mu\nu, \tilde{k}\tilde{k}'} 
e^{-i\varepsilon_{p}^{(\mu\tilde{k},\nu\tilde{k}')}
(t_c-t_p)}
\phi_{ \tilde{k},\tilde{k}'}^{(\mu,\nu)} 
\left\langle 0 \left|\hat{b}_{\beta, k'}\hat{b}_{\alpha, k}\hat{c}_{\mu, \tilde{k}}^\dag \hat{c}_{\nu, \tilde{k}'}^\dag \right|0\right\rangle,
\end{align}
where $\varepsilon_{p}^{(\mu\tilde{k},\nu\tilde{k}')}$ is the energy of the state $\hat{c}_{\mu, \tilde{k}}^\dag \hat{c}_{\nu, \tilde{k}'}^\dag |0\rangle$.
Substituting Eq.~(\ref{trans1}) into (\ref{trans2}), we obtain 
\begin{align}\label{bc2}
\psi_{ k,k'}^{(\alpha,\beta)}&=
\sum_{\mu\nu, \tilde{k}\tilde{k}'} 
e^{-i\varepsilon_{i}t_p}e^{-i\varepsilon_{p}^{(\mu\tilde{k},\nu\tilde{k}')}
(t_c-t_p)}\\\nonumber
&\times\left\langle 0 \left|\hat{b}_{\beta, k'}\hat{b}_{\alpha, k}\hat{c}_{\mu, \tilde{k}}^\dag \hat{c}_{\nu, \tilde{k}'}^\dag \right|0\right\rangle
\langle0|\hat{c}_{\nu, \tilde{k}'}\hat{c}_{\mu, \tilde{k}}|i\rangle\\\nonumber
&=e^{-i\varepsilon_{i}t_p}\langle0|\hat{b}_{\beta, k'}\hat{b}_{\alpha, k}|i\rangle
\sum_{\mu\nu, \tilde{k}\tilde{k}'}
e^{-i\varepsilon_{p}^{(\mu\tilde{k}, \nu\tilde{k}')}
(t_c-t_p)}
|\langle0|\hat{c}_{\nu, \tilde{k}'}\hat{c}_{\mu, \tilde{k}}|i\rangle|^2,
\end{align}
which indicates that the parameters of the intermediate time boundary do not disrupt the condition  $\psi_{k_1,k_2}^{(-,+)}=\pm\psi_{k_2,k_1}^{(-,+)}$ of Eq.~\eqref{bc1}.
That is, the suggested preparation method of Bell states is robust against the gradual variation of the time boundaries.

\section{Proof of entanglement robustness against environmental decoherence}\label{AppD}
To theoretically prove the entanglement robustness of the prepared states against environmental decoherence, the Lindblad master equation in momentum space is employed,
\begin{align}\label{lind_blade} 
\frac{d\hat{\rho}}{dt} = -i\left[\hat{\cal H},\ \hat{\rho}\right] + \gamma\sum_{\alpha,k} \left( \hat{a}_{\alpha,k}\hat{\rho}\hat{a}_{\alpha,k}^{\dag} - \frac{1}{2}\left\{ \hat{\rho}, \hat{a}_{\alpha,k}^{\dag}\hat{a}_{\alpha,k} \right\} \right),
\end{align} 
with the initial state taken as that in Eq.~\eqref{plwave}.
The Hamiltonian $\hat{\cal H}$ in Eq.~\eqref{Hamiltonian} is rewritten as
\begin{align}\label{Hamiltonian_i}
\hat{\cal H}=\sum_{\alpha,k}\varepsilon_{\alpha}(k)\hat{a}_{\alpha,k}^{\dag}
\hat{a}_{\alpha,k},
\end{align}
which is prior to the time boundary. 
Subsequent to the time boundary, we need to replace the $\hat{a}_{\alpha,k}^{\dag}$ ($\hat{a}_{\alpha,k}$) with $\hat{b}_{\alpha,k}^{\dag}$ ($\hat{b}_{\alpha,k}$).
To analyze the two-particle dynamics, we work in a truncated Fock space spanned by the vacuum, single-particle states, and two-particle states that involve only two momenta $k_1$ and $k_2$. This yields the following 9 basis states,
\begin{align}
  \begin{array}{ll}
    |1\rangle_i=|0\rangle, & |6\rangle_i=|-,k_1\rangle_i|-,k_2\rangle_i,  \\
    |2\rangle_i=|-,k_1\rangle_i, & |7\rangle_i=|-,k_1\rangle_i|+,k_2\rangle_i, \\
    |3\rangle_i=|-,k_2\rangle_i, & |8\rangle_i=|+,k_1\rangle_i|-,k_2\rangle_i,  \\
    |4\rangle_i=|+,k_1\rangle_i, & |9\rangle_i=|+,k_1\rangle_i|+,k_2\rangle_i, \\
    |5\rangle_i=|+,k_2\rangle_i,
  \end{array}\label{bs9a}
\end{align}
where the subscript $i$ denotes the time prior to $t_c$, that is, $|\alpha, k\rangle_i=\hat a_{\alpha, k}^\dag|0\rangle$.
The initial density matrix $\hat{\rho}_i=|i\rangle\langle i|$ has the form,
\begin{align}
\hat{\rho}_i=
\textrm{diag}(0,0,0,0,0,1,0,0,0).
\end{align}
The function
\begin{align}\label{time_evo_rho}
\hat{\rho}(t+\Delta t)=\hat{\rho}(t)+
\hat{R}(t)\cdot\Delta t
\end{align}
 with $\hat{R}(t)=d\hat{\rho}(t)/dt$ determines the evolution of $\hat{\rho}$ from $\hat{\rho}_i$.
 
Before the time boundary, only four elements of $\rho_{1,1}$, $\rho_{2,2}$, $\rho_{3,3}$ and $\rho_{6,6}$ evolve into nonzero terms, whereas all other elements remain at their initial values.
Accordingly, we can write $\hat{\rho}$ at time $t_{c}^-$ infinitesimally close to $t_c$ as
\begin{align}\label{rho_c-}
\hat{\rho}(t_{c}^-)
=\textrm{diag}\left(\rho_{11}, \rho_{22}, \rho_{33}, 0, 0, \rho_{66}, 0, 0, 0
\right)_{t=t_c^-}.
\end{align}
After the time boundary, the state bases turn into 
\begin{align}
  \begin{array}{ll}
    |1\rangle_f=|0\rangle, & |6\rangle_f=|-,k_1\rangle_f|-,k_2\rangle_f,  \\
    |2\rangle_f=|-,k_1\rangle_f, & |7\rangle_f=|-,k_1\rangle_f|+,k_2\rangle_f, \\
    |3\rangle_f=|-,k_2\rangle_f, & |8\rangle_f=|+,k_1\rangle_f|-,k_2\rangle_f,  \\
    |4\rangle_f=|+,k_1\rangle_f, & |9\rangle_f=|+,k_1\rangle_f|+,k_2\rangle_f, \\
    |5\rangle_f=|+,k_2\rangle_f,
  \end{array}\label{bs9b}
\end{align}
where the subscript $f$ denotes the time subsequent to $t_c$, that is, $|\alpha, k\rangle_f=\hat b_{\alpha, k}^\dag|0\rangle$.
The two state bases in Eqs.~\eqref{bs9a} and \eqref{bs9b} are connected by
\begin{align}\label{trans_rela1}
\left(\begin{matrix}
|1\rangle_f\\
|2\rangle_f\\
|3\rangle_f\\
|4\rangle_f\\
|5\rangle_f\\
|6\rangle_f\\
|7\rangle_f\\
|8\rangle_f\\
|9\rangle_f
\end{matrix}\right)=
\left(\begin{matrix}
1\\
& a_1 & & a_2\\
& & b_1 & & b_2\\
& c_1 & & c_2\\
& & d_1 & & d_2\\
& & & & & a_1b_1 & a_1b_2 & a_2b_1 & a_2b_2\\
& & & & & a_1d_1 & a_1d_2 & a_2d_1 & a_2d_2\\
& & & & & b_1c_1 & b_2c_1 & b_1c_2 & b_2c_2\\
& & & & & c_1d_1 & c_1d_2 & c_2d_1 & c_2d_2
\end{matrix}\right)
\left(\begin{matrix}
|1\rangle_i\\
|2\rangle_i\\
|3\rangle_i\\
|4\rangle_i\\
|5\rangle_i\\
|6\rangle_i\\
|7\rangle_i\\
|8\rangle_i\\
|9\rangle_i
\end{matrix}\right)
\end{align}
where $a$, $b$, $c$, and $d$ are non-zero constants and determined by Eqs.~\eqref{eig vec i} and \eqref{eig vec f}.
This transformation is responsible for the form of $\hat \rho$ at time $t_c^+$ infinitesimally after $t_c$, that is,
\setlength{\arraycolsep}{1.5pt}
\begin{align}\label{rho_c+}
\hat{\rho}(t_{c}^+)=
\left(
                \begin{matrix}
                  \rho_{1,1}  \\
                   & \rho_{2,2} &  &  \rho_{2,4}  \\
                   &  & \rho_{3,3} &  & \rho_{3,5}  \\
                   & \rho_{4,2} &  & \rho_{4,4}  \\
                   &  & \rho_{5,3} &  & \rho_{5,5}  \\
                   &  &  &  &  & \rho_{6,6} & \rho_{6,7} & \rho_{6,8} & \rho_{6,9} \\
                   &  &  &  &  & \rho_{7,6} & \rho_{7,7} & \rho_{7,8} & \rho_{7,9} \\
                   &  &  &  &  & \rho_{8,6} & \rho_{8,7} & \rho_{8,8} & \rho_{8,9} \\
                   &  &  &  &  & \rho_{9,6} & \rho_{9,7} & \rho_{9,8} & \rho_{9,9} \\
                \end{matrix}
              \right)_{t=t_c^+}.
\end{align}
In $\hat{\rho}(t_{c}^+)$, only non-zero elements are presented.
According to Eq.~\eqref{lind_blade}, $\hat \rho$ at $t>t_c$ has the form of $\hat{\rho}(t_{c}^+)$, as well as $\hat \rho_f$.

Since ${\hat \rho}_{\rm M}\ \left(\equiv \hat{Q}_{\rm M} {\hat \rho}_f \hat{Q}_{\rm M}^\dagger\right)$ involves only the state bases of $|7\rangle_f$ and $|8\rangle_f$, we focus on their elements in $\hat R(t)$ in Eq.~\eqref{time_evo_rho}, i.e., $R_{i,j}(t)$ with $i,j\in\{7, 8\}$.
According to Eq.~\eqref{lind_blade}, these elements have the form,
\begin{align}
R_{i,j}(t)=-2\gamma\rho_{i,j}(t),
\end{align}
and thus 
\begin{align}
{\hat \rho}_{\rm M}(t)={\hat \rho}_{\rm M}(t_c^+)e^{-2\gamma (t-t_c^+)}.\label{rhomt}
\end{align}
The equation \eqref{rhomt} tells that the entanglement of ${\hat \rho}_{\rm M}(t)$ is identical to that of ${\hat \rho}_{\rm M}(t_c^+)$.
Leveraging Eqs.~\eqref{trans_rela1} and \eqref{rho_c+}, one can find ${\hat \rho}_{\rm M}(t_{c}^+)$ as, 
\begin{align}
{\hat \rho}_{\rm M}(t_{c}^+)=
{1\over |a_1d_1|^2+|b_1c_1|^2}\left(
 \begin{array}{cc}
 |a_1d_1|^2 & a_1^{\ast}d_1^{\ast}b_1c_1 \\
 a_1d_1 b_1^{\ast}c_1^{\ast} & |b_1c_1|^2 \\
 \end{array}
 \right),
\end{align}\label{pure state}
which is normalized for convenience.
Since $\textrm{Tr}\left[{\hat \rho}_{\rm M}^2(t_c^+)\right]=1$, ${\hat \rho}_{\rm M}(t)$ describes a pure state.
Additionally, the condition in Eq.~\eqref{bellcondition}, implying $|a_1d_1|=|b_1c_1|$, leads to ${\hat \rho}_{\rm M}(t)$ representing a momentum-entangled Bell state.
These analyses are consistent with the numerical results in Fig.~\ref{Fig2}(e).

\section{Multi-band SSH model}\label{AppE}

\begin{figure*}[htbp]
  \centering
  \includegraphics[width=0.9\textwidth]{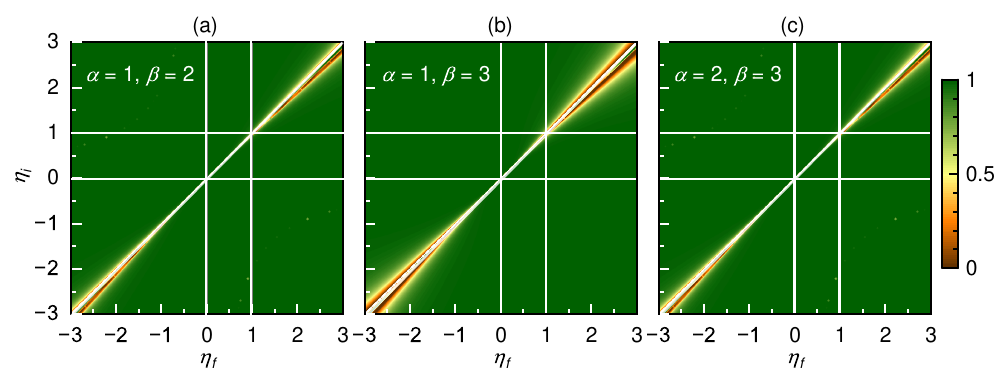}
  \caption{Variations of the von Neumann entropies for $|f_{\rm M}^{(\alpha,\beta)}\rangle$ with $\eta_i$ and $\eta_f$.
  Parameters: $k_0=0.75\pi$, $\Delta k=0.02\pi$.}\label{Fig6}
\end{figure*}

\subsection{Reflection and refraction angles}\label{angles}
Denoting the dispersion of the B$\alpha$ band in Fig.~\ref{Fig4}(a) as $\varepsilon_\alpha(k)$, we have its group velocity $v_\alpha=d\varepsilon_\alpha(k)/dk$.
The incident angle $\theta_i$ and scattering angles $\theta_\alpha$, marked in Fig.~\ref{Fig4}(b), are defined in a ``space-time" plane.
Using the way in Ref.~\cite{dong2024quantum}, we can find that they are determined by
\begin{align}
\tan{\theta_i}=v_i/v,~~~
\tan{\theta_\alpha}=v_\alpha/v,~~~\alpha\in\{1,2,\cdots,  M\},
\end{align}
where $v$ is the intra-cell coupling.
If we take the parameters in Fig.~\ref{Fig4}(b), i.e., $\eta_i=0.5$, $\eta_f=-1.1$, and $k_0=0.75\pi$, one has the three angles as $\theta_1=25.7^\circ$, $\theta_2=-33.4^\circ$, and $\theta_3=10.1^\circ$, where the minus sign represents the reflection.

\subsection{von Neumann entropy}\label{AppE2}

For the case with arbitrary $M$, the parametric conditions for preparing highly entangled states can be numerically explored. 
Similar to the $M = 2$ case, we start from the initial state  
\begin{align}  
|i\rangle = \hat{a}_{i,k_1}^\dagger \hat{a}_{i,k_2}^\dagger |0\rangle,  
\label{plwave3}  
\end{align}  
which represents a two-particle plane-wave pulse prepared on the \( \textrm{Bi} \) band, see Fig.~\ref{Fig4}(a). 
This state finally evolves into 
\begin{align}  
|f\rangle = \sum_{\mu\nu, kk'} \psi_{k,k'}^{(\mu,\nu)} \hat{b}_{\mu, k}^\dag \hat{b}_{\nu, k'}^\dag |0\rangle  
\end{align}  
where \( \mu, \nu \in \{1, 2, \dots, M\} \). 
Through projection with the operator \( \hat{Q}_{\rm M}^{(\alpha, \beta)} \), the post-projected state reads
\begin{align}\label{f12_M3}  
|f_{\rm M}^{(\alpha, \beta)}\rangle = \hat{Q}_{\rm M}^{(\alpha, \beta)}|f\rangle = \phi_{k_1,k_2}^{(\alpha, \beta)} |k_1\rangle_{\rm I} |k_2\rangle_{\rm II} + \phi_{k_2,k_1}^{(\alpha, \beta)} |k_2\rangle_{\rm I} |k_1\rangle_{\rm II},  
\end{align}  
where \( |k\rangle_{\rm I} = |\alpha, k\rangle \), \( |k'\rangle_{\rm II} = |\beta, k'\rangle \), and \( \phi_{k,k'}^{(\alpha, \beta)} = 2\psi_{k,k'}^{(\alpha,\beta)} \). 
The form of \( \psi_{k,k'}^{(\alpha,\beta)} \) is analogous to that in Eq.~(\ref{phi_ak1bk2}), provided that the parameters \( g \) and \( h \) are adapted to the \( M \)-band system. 
For \( M = 3 \), the entanglements \( S_{\alpha\beta} = -\left|\phi_{k_1,k_2}^{(\alpha,\beta)}\right|^2\log_2\left|\phi_{k_1,k_2}^{(\alpha,\beta)}\right|^2 - \left|\phi_{k_2,k_1}^{(\alpha,\beta)}\right|^2\log_2\left|\phi_{k_2,k_1}^{(\alpha,\beta)}\right|^2 \) are illustrated in Fig.~\ref{Fig6}. 
The three panels reveal a broad parametric range that can hold high entanglements between any two band branches. Similar to the case $M=2$, we also have \( \eta_i \neq \eta_f \) and \( \eta_i, \eta_f \notin \{0, 1\} \).


%

\end{document}